%% file: main.tex
\documentclass[acmsmall]{acmart}

\usepackage[utf8]{inputenc}
\usepackage[T1]{fontenc}

\usepackage{amsmath}
\usepackage{adjustbox}
\usepackage{booktabs}
\usepackage{enumitem}
\usepackage{balance}
\usepackage{float}
\usepackage{multirow}
\usepackage{listings}
\usepackage[font=small]{subfig}
\usepackage[absolute]{textpos}
\usepackage{tcolorbox}

\setcopyright{acmlicensed}
\acmJournal{PACMNET}
\acmYear{2023} \acmVolume{1} \acmNumber{CoNEXT2}
\acmArticle{6} \acmMonth{9} \acmPrice{15.00}
\acmDOI{10.1145/3609423}
\received{November 2022}
\received[revised]{May 2023}
\received[accepted]{June 2023}

\usepackage{tikz}
\usetikzlibrary{
  arrows,
  arrows.meta,
  calc,
  colorbrewer,
  decorations.pathreplacing,
  calligraphy,
  external,
  fit,
  matrix,
  patterns,
  positioning,
  shadows,
  shapes,
  shapes.misc,
}
\pgfdeclarelayer{bg}    
\pgfdeclarelayer{fg}    
\pgfsetlayers{bg,main,fg}  
\tikzexternaldisable%

\hypersetup{
  breaklinks,
}

\graphicspath{{figs/}}

\input{commands}

\title{Securing Name Resolution in the IoT\@: DNS over CoAP}

\author{Martine S. Lenders}
\orcid{0000-0001-7378-1045}  
\affiliation{%
  \institution{Freie Universit\"at Berlin}
  \city{Berlin}
  \country{Germany}
}
\affiliation{%
  \institution{TU Dresden}
  \city{Dresden}
  \country{Germany}
}
\email{m.lenders@fu-berlin.de}

\author{Christian Ams\"uss}
\orcid{0000-0001-5168-5861}  
\affiliation{
  \institution{Unaffiliated}
  \city{Vienna}
  \country{Austria}
}
\email{christian@amsuess.com}

\author{Cenk G{\"u}ndogan}
\orcid{0000-0002-9395-625X}  
\affiliation{%
  \institution{Huawei Technologies}
  \city{Munich}
  \country{Germany}
}
\email{cenk.gundogan@huawei.com}

\author{Marcin Nawrocki}
\orcid{0000-0002-6308-5502}  
\affiliation{%
  \institution{Freie Universit\"at Berlin}
  \city{Berlin}
  \country{Germany}
}
\affiliation{%
  \institution{TU Dresden}
  \city{Dresden}
  \country{Germany}
}
\email{marcin.nawrocki@fu-berlin.de}

\author{Thomas C. Schmidt}
\orcid{0000-0002-0956-7885}  
\affiliation{%
  \institution{HAW Hamburg}
  \city{Hamburg}
  \country{Germany}
}
\email{t.schmidt@haw-hamburg.de}

\author{Matthias W\"ahlisch}
\orcid{0000-0002-3825-2807}  
\affiliation{%
  \institution{TU Dresden}
  \city{Dresden}
  \country{Germany}
}
\affiliation{%
  \institution{Barkhausen Institut}
  \city{Dresden}
  \country{Germany}
}
\email{m.waehlisch@tu-dresden.de}

\renewcommand{\shortauthors}{Lenders et al.}

\begin{document}

\input{abstract}



\maketitle

\begin{textblock}{13.3}(1.4,0.6)
	\noindent
  \begin{tcolorbox}[left=1pt,right=1pt,top=1pt,bottom=1pt,colframe=black, boxrule=1pt]
    \footnotesize
    If you refer to this paper, please cite the peer-reviewed publication:
    M.S. Lenders, C. Amsüss, C. Gündogan, M. Nawrocki, T.C. Schmidt, and M. Wählisch. 2023.
    Securing Name Resolution in the IoT: DNS over CoAP. \textit{Proceedings of the ACM on
    Networking (PACMNET)} 1, CoNEXT2, Article 6 (September 2023), 25 pages.
  \end{tcolorbox}
\end{textblock}

\input{tex/intro}
\input{tex/related-work}
\input{tex/dns-empiric}
\input{tex/doc}
\input{tex/evaluation-base}
\input{tex/evaluation-caching}
\input{tex/discussion}
\input{tex/conclusion}
\input{tex/acks.tex}

\label{lastpage}  

\bibliographystyle{ACM-Reference-Format}

\bibliography{./local,own,rfcs,ids,iot,layer2,security}

\appendix
\input{tex/appendix_blockwise}
\input{tex/appendix_impl}
\input{tex/appendix_riot_config}
\input{tex/appendix_coap_pkt_size}
\input{tex/acronyms.tex}

\end{document}

%% file: commands.tex
\usepackage{pifont}
\newcommand{\cmark}{\ding{51}}%
\newcommand{\xmark}{\ding{56}}%

\usepackage{fontawesome5}
\usepackage{bm}

\usepackage{xspace}
\newcommand{\eg}{\textit{e.g.,}~}
\newcommand{\ie}{\textit{i.e.,}~}

\newcommand{\one}{({\em i})\xspace}
\newcommand{\two}{({\em ii})\xspace}
\newcommand{\three}{({\em iii})\xspace}
\newcommand{\four}{({\em iv})\xspace}
\newcommand{\five}{({\em v})\xspace}

\usepackage{cleveref}
\crefname{section}{Section}{Sections}
\crefname{figure}{Figure}{Figures}
\crefname{table}{Table}{Tables}
\crefname{equation}{Equation}{Equations}

\makeatletter
\renewcommand{\paragraph}[1]{\noindent{\bf #1}\hspace{0.25ex \@plus1ex \@minus.2ex}}
\makeatother

\definecolor{darkblue}{HTML}{330099}
\definecolor{C0}{HTML}{d7191c}
\definecolor{C1}{HTML}{fdae61}
\definecolor{C2}{HTML}{abdda4}
\definecolor{C3}{HTML}{2b83ba}
\definecolor{C4}{HTML}{3d194a}
\colorlet{C-udp}{C0}
\colorlet{C-dtls}{C1}
\colorlet{C-oscore}{C2}
\colorlet{C-coaps}{C3}
\colorlet{C-coap}{C4}

\newcommand{\timeconstraint}[6][0]{%
  \node [coordinate] (#2 start) at (t-start |- #3.north) {};
  \node [coordinate] (#2 end) at (t-start |- #4.north) {};
  \draw [densely dashed, draw=gray!75!black] (#3.north) -- (#2 start);  
  \draw [densely dashed, draw=gray!75!black] (#4.north) -- (#2 end);  
  \draw [draw=gray, rounded corners=1mm] ($(#2 start)$) -- ($(#2 start) - (1mm + #1, 0)$) -- ($(#2 end) - (1mm + #1, 0)$) -- (#2 end);  

  \ifx&#6&%
    \node [anchor=east, xshift=0.7mm, font=\scriptsize\color{gray!75!black}] (#2) at ($(#2 start)!0.5!(#2 end) - (1mm + #1, 0)$) {#5};%
  \else%
    \node [anchor=east, xshift=0.7mm, font=\scriptsize\color{gray!75!black}, yshift={#6}] (#2) at ($(#2 start)!0.5!(#2 end) - (1mm + #1, 0)$) {#5};%
  \fi%
}

\newlength{\aepaperxadjust}
\newlength{\aepaperyadjust}


%% file: abstract.tex
\begin{abstract}
In this paper, we present the design, implementation, and analysis of DNS over CoAP~(DoC), a new proposal for secure and privacy-friendly name resolution of constrained IoT devices.
We implement different design choices of DoC in RIOT, an open-source operating system for the IoT, evaluate performance measures in a testbed, compare with DNS over UDP and DNS over DTLS, and validate our protocol design based on empirical DNS IoT data.
Our findings indicate that plain DoC is on par with common DNS solutions for the constrained IoT but significantly outperforms when additional standard features of CoAP are used such as caching.
With OSCORE, we can save more than 10 kBytes of code memory compared to DTLS, when a CoAP application is already present, and retain the end-to-end trust chain with intermediate proxies, while leveraging features such as group communication or encrypted en-route caching.
We also discuss a compression scheme for very restricted links that reduces data by up to 70\%.
\end{abstract}


%% file: tex/intro.tex
\section{Introduction}\label{sec:intro}

\begin{table}
  \centering
  \caption{Comparison of DNS transport features. The contributions of this paper are highlighted in bold.}%
  \label{tab:feature-comparison}
  \footnotesize
  \renewcommand{\r}[1]{#1}
  \begin{tabular}{rccccccccccc}
    \toprule
                                    & \multicolumn{9}{c}{DNS over} \\
                                      \cmidrule{2-10}
    Protocol Feature                & \r{UDP} & \r{TCP} & \r{DTLS}& \r{TLS} & \r{QUIC}&\r{HTTPS}& \r{\bf CoAP}&\r{\bf CoAPS}& \r{\bf OSCORE}\\
    \midrule
    Message Segmentation            & \xmark  & \cmark  & \xmark  & \cmark  & \cmark  & \cmark  & \cmark  & \cmark  & \cmark \\
    Message Authentication          & \cmark  & \cmark  & \cmark  & \cmark  & \cmark  & \cmark  & \cmark  & \cmark  & \cmark \\
    Message Encryption              & \xmark  & \xmark  & \cmark  & \cmark  & \cmark  & \cmark  & \xmark  & \cmark  & \cmark \\
    Message Format Multiplexing     & \xmark  & \xmark  & \xmark  & \xmark  & \xmark  & \cmark  & \cmark  & \cmark  & \cmark \\
    Shares protocol with application & \xmark  & \xmark  & \xmark  & \xmark  & \xmark  & \cmark  & \cmark  & \cmark  & \cmark \\
    Suitability for Constrained IoT & \cmark  & \xmark  & \cmark  & \xmark  & \xmark  & \xmark  & \cmark  & \cmark  & \cmark \\
    Content Secure En-route Caching & \xmark  & \xmark  & \xmark  & \xmark  & \xmark  & \xmark  & \xmark  & \xmark  & \cmark \\
    \bottomrule
  \end{tabular}
\end{table}

The Internet of Things~(IoT) deployment extends from simple environmental sensors, industrial monitoring and control to various consumer-grade products, such as home cameras and Smart-TVs.
Most IoT operations require frequent access to data or (cloud-)services, commonly addressed by names.  
Unprotected name resolution from such devices---often in long range undisclosed radio networks---raises concerns regarding security and privacy, as names may carry contextual semantics that enable fingerprinting for third-parties.
The system-wide implementation of a vulnerable protocol such as DNS over UDP also opens IoT deployments to large-scale botnet creation.
Hence, protecting name resolution in the IoT plays a central role.

IoT devices are often constrained.
These \textit{things} commonly interconnect wirelessly and remain independent of the power grid---they typically operate on batteries or harvest energy from the environment.
Hardware platforms are kept simple to prolong operation times and reduce unit costs.
Even the most powerful devices according to the common IETF classification~\cite{RFC-7228} show orders of magnitude less memory than general-purpose hardware platforms (see \cref{tab:deviceclasses}).
These devices still require protected name resolution to find, \eg cloud services~\cite{hkccs-diorc-20,smsgf-ddiib-22}, possibly via long-range radio communication through untrusted gateways, \eg~\cite{ttn,lksw-saccl-22}.

Protecting the name resolution of DNS strengthens privacy and security~\cite{zhhwm-cdips-15}.
Common uses of DNS on top of encrypted transport, though, \eg DNS over HTTPS~(DoH)~\cite{RFC-8484}, DNS over TLS~(DoT)~\cite{RFC-7858}, or DNS over QUIC~(DoQ)~\cite{RFC-9250}, conflict with low hardware resources of constrained class~1 or class~2 devices.
DNS over DTLS~(DoDTLS)~\cite{RFC-8093} does not provide means for message segmentation as imposed by the small link layer frame sizes of constrained networks (see~\cref{tab:network-constraints}).

In this paper, we present DNS over CoAP~(DoC), a secure and privacy-friendly DNS resolution protocol for the constrained IoT\@.
The Constrained Application Protocol~(CoAP)~\cite{RFC-7252} was standardized by the IETF
as a lightweight IoT alternative to HTTP and is widely available.
CoAP is based on UDP but provides transactional message contexts, reliability retransmission functions, and en-route caching on dedicated forward proxies.
Security extensions either use Datagram Transport Layer Security~(DTLS)~\cite{RFC-8094} or Content Object Security for Constrained RESTful Environments~(\mbox{OSCORE})~\cite{RFC-8613}. DoC leverages these security extensions to query the DNS privately, securely, and yet efficiently enough to comply with the low-end IoT\@.

\paragraph{Challenges:} Designing and implementing DNS for the constrained IoT is more challenging than DoH when embedded into common IETF~protocols.
\begin{enumerate}[itemsep=0pt, topsep=0pt, parsep=0pt, leftmargin=*]
  \item Common DNS answers are large and lead to packet fragmentation, which should be avoided on lossy IoT links~\cite{lsw-ffln-21}.
  \item Recently released CoAP methods offer differing features and tradeoffs.
  \item En-route caching has more importance in CoAP to decouple lossy IoT links from content delivery~\cite{gasw-triwt-20} and fragmentation avoidance benefits from the cache validation mechanism of CoAP~\cite{RFC-7252}.
    This extends the design space beyond the purely client-based DoH caching.
  \end{enumerate}

\paragraph{Contributions:} A feature comparison of the different DNS transports, which highlights the achievements of our work, is shown in \cref{tab:feature-comparison}.
The main contributions of this paper are as follows.

\begin{enumerate}[itemsep=0pt, topsep=0pt, parsep=0pt, leftmargin=*]
  \item We analyze the impact of IoT~naming on the design of DNS resolution based on an empirical data set of characteristic IoT~domain names. We compare with name requests observed at a large regional Internet eXchange Point~(IXP).
  (\cref{sec:dns-empiric})

  \item We design the DNS over CoAP~(DoC), which efficiently embeds the DNS semantics into
	   the rich feature set of the CoAP protocol suite to enable end-to-end protection, block-wise transfer, group communication, caching, as well as an option for compression.
    (\cref{sec:doc})

  \item A system-level analysis conducted on real IoT hardware using key properties gathered in \cref{sec:dns-empiric} reveals that DoC performance is at least on par with generic UDP-based DNS transport. Additional features increase the DoC performance further.
    (\cref{sec:eval-base,sec:eval-caching})

  \item We discuss the utility of a potential new media type to transport DNS messages over DoC or DoH\@. (\cref{sec:discussion})
\end{enumerate}

\begin{table}
  \centering
  \caption{Constraints of DoC~target platforms.}%
  \label{tab:constraints}
  \setlength{\tabcolsep}{4.0pt}
  \footnotesize
  \subfloat[Memory constraints~\cite{RFC-7228}]{%
    \label{tab:deviceclasses}
    \begin{tabular}{rrrr}
      \toprule
      Memory & \multicolumn{1}{c}{Class 0} & \multicolumn{1}{c}{Class 1} & \multicolumn{1}{c}{Class 2}\\
      \midrule
      RAM [kBytes] & $\ll$ 10 & $\approx$ 10 & $\approx$ 50\\[0.25em]
      ROM [kBytes] & $\ll$ 100 & $\approx$ 100 & $\approx$ 250\\
      \bottomrule
    \end{tabular}
  }
  \hfill
  \subfloat[Network constraints~\cite{RFC-8376, IEEE-802.15.4-16, la-lrp-22, RFC-7668, lmmddd-mtmd802154-05}]{%
    \label{tab:network-constraints}
    \begin{tabular}{rrrrrr}
      \toprule
      Characteristic        & IEEE 802.15.4 & BLE         & LoRaWAN       & NB-IoT \\
      \midrule
      Data rate [kBit/s]    & 124--162      & 125--2000   & 0.3--5        & 30--60 \\[0.25em]
      Frame size [bytes]    & 127           & $\geq$ 1280 & 59--250       & 1600 \\
      \bottomrule
    \end{tabular}
  }
  \vskip-1em
\end{table}

The remainder of this paper is structured as follows.
After exploring the problem space and the related work in \cref{sec:rw}, we present our contributions in \crefrange{sec:dns-empiric}{sec:eval-caching}, discuss further potential for DoC in \cref{sec:discussion}, and conclude with an outlook in \cref{sec:conclusion}.


%% file: tex/related-work.tex
\section{The Problem of Name Resolution in the IoT and Related Work}\label{sec:rw}

\subsection{The Need for Secure Name Resolution}

\paragraph{DNS resolvers on IoT devices.}
The DNS serves as an indirection to reach IP~endpoints, enabled by a DNS (stub) resolver at Internet nodes.
Omitting the resolver on the IoT device would introduce several drawbacks.
First, IP~addresses to access Internet services, \eg cloud backends, need to be preconfigured on the IoT device.
This reduces flexibility and adds additional burden, if preconfigured addresses need changing.
Second, offloading the name resolution to more powerful border routers (or gateways) would require application-specific deep packet inspection on the gateway or an additional name resolution protocol between gateway and IoT device.
We argue that a gateway should remain as transparent as possible and that DNS resolution should be interoperable throughout the global Internet.
Finally, we emphasize that unprotected DNS over UDP is already available in popular IoT operating systems, such as ARM~Mbed~\cite{mbed-6.16-dns}, RIOT~\cite{riot-dns}, and Zephyr~\cite{zephyr-3.2-dns}.

\paragraph{Threats to end user privacy.}
Meta-data about communication can leak sensitive information such as sleeping habits~\cite{axrnf-kshpats-19}.
One such meta-data are the hostnames IoT devices resolve~\cite{rdcmkh2019moniotr,ppaa2020iotfinder,smgsf-ddibe-22}: Names provide application- or service-specific information.
Plain DNS~queries concurrent to (protected) application traffic may disclose the context of confidential communication, reveal behavioral patterns, or uncover hints for fingerprinting victims~\cite{lllhd-elmdh-19}.
In this paper, we close the gap in privacy-friendly DNS resolution~\cite{lgs-sdnsenc-22} by designing and analyzing a lightweight protocol that makes use of emerging IoT~standards to obfuscate DNS queries and responses.

\paragraph{Threats to infrastructure security.}
Leaked information, may it be personal or not, is a security risk by itself. An attacker can infer knowledge about the network or applications from it and plan attacks accordingly.
The IoT, however, repeatedly takes center stage in large-scale distributed Denial of Service~(DDoS) attacks against the Internet infrastructure.
As an example, the Mirai botnet caused 600 GBytes/s of incoming traffic, taking down commercial servers in 2016~\cite{aabbb-umb-17,gh-didi-20}.

DNS rebinding attacks are ways to redirect a requester to malicious sites, which then can deploy malware on a device~\cite{l-dra-19,t-cssst-22,r-ahrnp-14,hkccs-diorc-20}.
Securing the name resolution with encryption helps mitigate such attacks and provides another factor of name authentication directly on the end device~\cite{h-srtarsa-19}.

\subsection{Challenges from the Constrained IoT}
The use of DNS over TCP~\cite{RFC-7766} mitigates DNS-based DDoS attacks, but has limited support by many resolvers~\cite{mrs-asdw-22}.
TCP and its TLS extension introduce an increased transport complexity. Even though TCP can be deployed in some constrained scenarios~\cite{kakc-ptlpwn-20}, it is not suitable for very restricted link-layer technologies that exhibit small packet sizes and long duty-cycles, such as LoRaWAN~\cite{RFC-9006}.
DoT~\cite{RFC-7858}, DoH~\cite{RFC-8484}, and most recently DoQ~\cite{RFC-9250} are mechanisms to protect the confidentiality and integrity of DNS traffic on the Internet.
They employ transport layer security and maintain session state between two endpoints, which prevents IP spoofing.
The first two approaches build on TCP and their performance significantly drops when network conditions degrade~\cite{hbshf-ceddd-20}.
This property conflicts with constrained networks where links commonly saturate.
DoQ uses UDP, but despite its performance advantages over DoT and DoH~\cite{kdgb-ortaf-22}, deployment in low-power regimes is challenging due to its use of TLS~\cite{RFC-8446}.

DoDTLS~\cite{RFC-8094} is an alternative that also runs on datagram transport\@.
In addition to reduced protocol complexity compared to the former approaches, this transport does not suffer from head-of-line blocking, which frequently occurs in constrained networks.
Nevertheless, DoDTLS faces issues with larger messages exceeding the Path Maximum Transmission Unit~(PMTU)~\cite{RFC-8094} and forces applications into fragmentation.
Moreover, IoT link layers, such as IEEE~802.15.4, LoRaWAN, or DSME-LoRa, often support only a few hundred bytes~\cite{IEEE-802.15.4-16,la-lrp-19,aksw-dslrc-22}.
These limits (see \autoref{tab:network-constraints}) are easily reached by rather small DNS~queries or responses as we will show later.
Even though adaption layers for IPv6, such as 6LoWPAN~\cite{RFC-4944} or Static Context Header Compression and Fragmentation~(SCHC)~\cite{RFC-9011}, offer fragmentation between the link and network layer, they introduce higher packet loss and latency~\cite{lsw-ffln-21}.

Despite the advantages of DoDTLS, there are certain drawbacks with protecting the transport in the IoT use case~\cite{gasw-cosit-22}.
\one~IoT networks may connect to gateways that bridge between transports, \eg between UDP and TCP\@.
This burdens the end-to-end protection, since gateways need to be included in trust relationships to re-encrypt the data between endpoints.
\two A loose coupling and caching are favored techniques in the IoT to deal with mobility and network partitioning~\cite{gasw-triwt-20}, but the established security sessions are deeply rooted on the transport and harden the endpoint paradigm.
\three Long duty-cycles in lossy, constrained networks as in LoRaWAN conflict with the handshake requirements of DTLS\@.

While current standardization efforts for DTLS, \eg Connection Identifiers~\cite{RFC-9146}, help to address the drawbacks, there is another undertaking in the IETF CoRE working group to provide secure communication.
OSCORE~\cite{RFC-8613} protects messages with encryption on the object-level instead of the transport-level.
It fully integrates with the CoAP~\cite{RFC-7252} ecosystem, ensures end-to-end protection across gateways, allows for a protected multiparty communication, and makes encrypted and authenticated CoAP messages cacheable on untrusted proxies.
In addition, OSCORE provides better performance compared to CoAP over DTLS~\cite{gbpst-eposcore-21}.

CoAP was designed as the HTTP for the IoT, and thus is the candidate for protective measures analog to DoH\@.
The CoAP ecosystem facilitates power-efficient, privacy-friendly DNS queries in the IoT and mitigates the size of DDoS attacks through bandwidth reduction and peer authentication.


%% file: tex/dns-empiric.tex
\section{Empirical View on IoT DNS Traffic}\label{sec:dns-empiric}

In this section, we motivate our design decisions.
To this end, we empirically analyze DNS traffic produced by end-consumer IoT devices and compare to flow samples from a large regional European~IXP\@.

\subsection{Data Corpus}
Identifying IoT-specific traffic is challenging.
We rely on data from three common projects, Your\-Things~\cite{alam2019yourthings}, IoTFinder~\cite{ppaa2020iotfinder}, and MonIoTr~\cite{rdcmkh2019moniotr}, all collected throughout 2019, which captured and annotated IoT~traffic based on ground truth.
YourThings and IoTFinder also provide traffic from desktop computers, phones, tablets, gaming consoles, and Wi-Fi access points.
We exclude such traffic.
All three IoT data sets include both unicast DNS and multicast DNS~(mDNS)~\cite{RFC-6762} traffic.
As mDNS is integral to DNS Service Discovery (DNS-SD), which is used in many use cases for IoT device discovery, we keep mDNS traffic.
Overall, the aggregation of all three data sources include data from over 90 consumer-grade IoT devices by more than 50 manufacturers and contains 0.2~million DNS and mDNS queries and 1.3~million corresponding responses in total for a total of 2336 unique names.
The IoTFinder data only contains responses but we can infer the queries from their question section.

To compare IoT-specific DNS traffic with DNS traffic from common Internet devices, we leverage sFlow~\cite{RFC-3176} samples collected in January 2022 at a European IXP\@.
Our IXP data contains 1.6~million unicast DNS queries and 2.4~million responses and is based on a sampling rate of 1/16000 packets.
Packets are truncated to 128~bytes.
For privacy regulation compliance, we strip all names and replace them with the target analysis data (\eg name lengths) before exporting them for our analysis.
As such, we were not able to count the total number of unique names.
Before stripping, we confirmed that no side effects on the name lengths were introduced due to query name minimization~\cite{RFC-7816}.

\begin{table}
  \centering
  \caption{Statistical key properties of domain names queried by IoT devices compared to domain names visible at an IXP\@. $\mu$ denotes the mean, $\sigma$ the standard deviation, $Q_1$ the first quartile, $Q_2$ the second quartile (or median), and $Q_3$ the third quartile.}%
  \label{tab:iot-data-name-stats}
  \footnotesize
  \setlength{\tabcolsep}{1.1em}
  \begin{tabular}{rccccrrccc}
    \toprule
    & & \multicolumn{8}{c}{Lengths of domain names [chars]} \\
    \cmidrule(l){3-10}
    Data source & Unique names [\#]  & min & max & mode & \multicolumn{1}{c}{$\mu$} & \multicolumn{1}{c}{$\sigma$} & $Q_1$ & $Q_2$ & $Q_3$ \\
    \midrule
    YourThings~\cite{alam2019yourthings} & 1293 & 2 & 83 & 31 & 24.5 & 9.7 & 18 & 24 & 30 \\
    IoTFinder~\cite{ppaa2020iotfinder} & 1097 & 7 & 82 & 24 & 26.8 & 10.5 & 20 & 24 & 30 \\
    MonIoTr~\cite{rdcmkh2019moniotr} & 695 & 9 & 83 & 18 & 27.1 & 14.7 & 18 & 23 & 30 \\
    \cmidrule(l){2-10}
    IoT total & 2336 & 2 & 83 & 24 & 25.9 & 11.3 & 19 & 24 & 30 \\
    \midrule
    IXP & --- & 0 & 68 & 17 & 26.1 & 11.7 & 17 & 25 & 33 \\
    \bottomrule
  \end{tabular}
\end{table}

\begin{figure}
  \centering
  \subfloat[IoT devices]{
    \centering
    \includegraphics{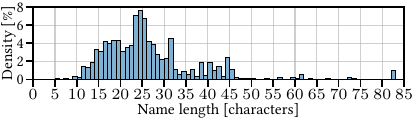}\label{fig:iot-data-name-len-iot}%
  }
  \subfloat[Internet devices]{
    \centering
    \adjustbox{clip=True, trim=1.4em 0 0 0}{%
      \includegraphics{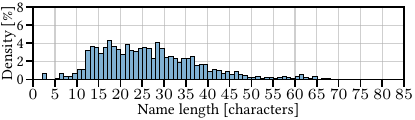}%
    }\label{fig:iot-data-name-len-ixp}
  }
  \setlength{\belowcaptionskip}{-0.9em}
  \caption{Distribution of name lengths; names queried by different devices connected via the Internet.}%
  \label{fig:iot-data-name-len}
\end{figure}

\subsection{Results}
\paragraph{How long are names requested by IoT devices?}
\cref{fig:iot-data-name-len} shows a normalized histogram over all queried names seen, for both the IoT data sets (\cref{fig:iot-data-name-len-iot}) and the IXP data set (\cref{fig:iot-data-name-len-ixp}).
The statistical key properties for the name lengths of each data set and the aggregated IoT data set are shown in \cref{tab:iot-data-name-stats}.

The median of the name lengths is 23 or 24~characters, depending on the IoT data set, which is similar to the median of 25~characters in the IXP data set (see $Q_2$ in \cref{tab:iot-data-name-stats}).
Many cloud and CDN names, such as \verb$e123.abcd.akamaiedge.net$ (name modified for sake of privacy), gather around this name length.
Significantly longer names are used for certain mDNS applications, \eg for reverse DNS or to identify local devices via a UUID\@.
As such, we do not see these longer name lengths at the IXP\@.

Combining all IoT data sets (IoT total), we find a median of 24~characters for the domain names, which results into 18.8\% of 127~bytes of the IEEE 802.15.4 link layer PDU\@.
Considering the mean of 25.9~characters, even more space of the link layer PDU is occupied by the name itself.
LoRaWAN reduces the PDU to only 59 bytes.
In this case, the name would require 40.7\% of the available space.

\paragraph{What kind of records are requested?}
\cref{tab:iot-data-rr} presents the relative ratio of the queried record types available in the \texttt{IN} class based on our data sets.
\texttt{A} records are in all data sets the most requested records, with \texttt{AAAA} records being close second.
With growing deployment of IPv6, these numbers will change in favor of more \texttt{AAAA} queries.
When not accounting for mDNS, these are $>$99\% of all records in the IoT\@.
With mDNS we also see more records associated with service discovery, namely \texttt{ANY}, \texttt{PTR}, \texttt{SRV}, and \texttt{TXT} records~\cite{RFC-6762,RFC-6763}.

We mostly see \texttt{A} and \texttt{AAAA} records at the IXP, in addition to several rarely requested resource records (3.5\% in total),
The largest remaining part are \texttt{PTR} and \texttt{HTTPS}~\cite{draft-ietf-dnsop-svcb-https} records.

\paragraph{Other DNS data of interest.}
We also analyzed the number of entries in each DNS~section and the overall response lengths.
In response sections, we find events that include more than 255~entries---the overflow point of the count fields into 2-byte numbers---but the percentage is low and mostly relate to mDNS\@.
These large numbers originate from unrequested \texttt{NS} records that advertise name servers in the authority section and the associated \texttt{A} or \texttt{AAAA} records for these advertised name servers in the additional section.
Providing these optional data seems to be common practice by cloud and CDN providers.
The question section, on the other hand, always contains only 1~entry.
This complies with common resolver behavior to ignore queries or cause an error when they include a question section with more than 1~entry~\cite{RFC-1035}.

Responses can become very long, containing 400 to 600 bytes, in certain cases even more than 1 kByte, even for the IoT devices.
This is caused by the long authority and additional sections as described above.

\begin{table}
  \centering
  \caption{Queried record types in \texttt{IN} class.}%
  \label{tab:iot-data-rr}
  \setlength{\tabcolsep}{0.8em}
  \footnotesize
  \begin{tabular}{lccccccccc}
    \toprule
    & \multicolumn{9}{c}{Record type}
    \\
    \cmidrule{2-10}
    Data set         & \texttt{A}  & \texttt{AAAA} & \texttt{ANY}  & \texttt{HTTPS}  & \texttt{NS} & \texttt{PTR}  & \texttt{SRV}  & \texttt{TXT}  & \texttt{Other}  \\
    \midrule
    IoT   w/ mDNS  & 53.6\%      & 16.4\%        & 8.2\%         & ---             & ---         & 19.6\%        & 1.0\%         & 1.2\%         & $<$0.1\%        \\
                    IoT w/o mDNS  & 75.8\%      & 23.5\%        & ---           & ---             & ---         & 0.3\%         & ---           & 0.1\%         & 0.3\%           \\
    IXP                 & 64.5\%      & 17.6\%        & 1.7\%         & 9.1\%           & 0.7\%       & 1.8\%         & 0.4\%         & 0.7\%         & 3.5\%           \\
    \bottomrule
  \end{tabular}
\end{table}

\paragraph{Lessons learned.}
Our analysis revealed that long names with a median of 24 characters are common, even in IoT scenarios.
Without a dedicated DNS message compression this can lead to fragmentation in constrained IoT scenarios.
While not part of the main design of DoC, the Content-Format option of CoAP offers the opportunity to compress the message format, which will lead to much smaller message sizes than the DNS wire format can provide.

IoT devices mainly query \texttt{A} and \texttt{AAAA} resource records, except when they use DNS-SD\@.
Service discovery implemented in DNS-SD often involves other resource records.
Group OSCORE~\cite{draft-ietf-core-oscore-groupcomm} may offer an encrypted but lightweight solution for multicast DNS-SD\@.

In constrained IoT use cases, the authority and additional sections must only be provided if necessary.
Unsolicited \texttt{NS} records serve little purpose in a constrained environment and should be omitted.


%% file: tex/doc.tex
\begin{figure}
  \centering
  \includegraphics{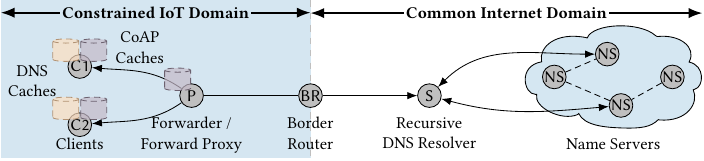}
  \caption{Typical deployment of the DNS over CoAP~(DoC) architecture. DoC protects name resolution between the clients running on IoT devices and the recursive DNS resolver.}%
  \label{fig:architecture}
\end{figure}

\section{Design of DNS over CoAP~(DoC)}\label{sec:doc}

In this section, we define DNS over CoAP~(DoC), a protocol to query the Domain Name System~(DNS) and retrieve responses over the Constrained Application Protocol~(CoAP)~\cite{RFC-7252} (see \cref{fig:architecture}).
This protocol work is also in the standardization process of the IETF~\cite{draft-ietf-core-dns-over-coap}.
The goal is to ensure message integrity and confidentiality by mapping each DNS query-response pair to a CoAP message exchange, secured on the transport via DTLS~\cite{RFC-6347} or on the object level via OSCORE~\cite{RFC-8613}.
DoC itself forms a thin layer between CoAP and DNS, comparable to DoH~\cite{RFC-8484} which maps DNS message pairs into HTTP message pairs.

Using CoAP for DNS resolution provides the following advantages for the constrained IoT\@:
\one CoAP runs on UDP, which
 does not introduce additional connection setup nor state.
CoAP provides optional reliability using a simple mechanism of acknowledgments and retransmissions.
\two Proxy operations and en-route response caches decouple packet loss from content delivery, which strengthens wireless networking~\cite{gasw-triwt-20}.
\three The Content-Format option of CoAP provides potential for future, compressed DNS messages to reduce fragmentation on the link layer.
\four CoAP provides block-wise transfer to fragment and reassemble large messages on the application layer (see \cref{sec:block-wise}), which are likely in DNS as seen in \cref{sec:dns-empiric}.

\subsection{Protocol Overview}
The DoC message exchange requires a mapping of DNS queries to CoAP requests and of DNS responses to CoAP replies.

\paragraph{Request mapping.}
A DoC client can send a DNS query over CoAP by embedding the DNS wire format of a DNS query into a CoAP message using either GET, POST, or FETCH~\cite{RFC-8132} requests, each of them provide different features.

GET and POST methods are already supported by DoH and their behavior could be translated to DoC\@:
Using GET, the DNS query requires encoding within the request URI\@.
As such, a DoC resource needs to be configured as a \emph{URI template}~\cite{RFC-6570}, describing the position of the DNS query in the URI as a variable.
GET allows for caching of subsequent responses but prevents block-wise transfer and demands a \emph{URI template processor} for resolving the DNS query variable at the constrained client side.
POST, on the other hand, carries the DNS query in the CoAP body, reducing the complexity of an additional URI template processor.
As a drawback, though, it does not allow for caching since the payload of the request is not taken into account for a cache key.
To allow for both caching and block-wise transfer, a DoC client can use FETCH~\cite{RFC-8132}.
FETCH is currently not supported by all CoAP~implementations, but extending them is trivial.
As such, compared to DoH, where GET and POST are used primarily, FETCH is the preferred method for DoC\@.
\cref{tab:doc-methods-comp} displays the different benefits and drawbacks of the three CoAP~methods.

\paragraph{Response mapping.}
A DoC~server sends a DNS response over CoAP by encoding the wire format of the DNS response in the payload of a CoAP response.

\begin{table}
  \centering
  \caption{Comparison of request methods considered for DoC\@.}%
  \label{tab:doc-methods-comp}
  \footnotesize
  \begin{tabular}{p{5cm}ccc}
    \toprule
    Feature & GET     & POST    & FETCH \\
    \midrule
    Cacheable                         & \cmark  & \xmark  & \cmark \\
    Application data carried in body  & \xmark  & \cmark  & \cmark \\
    Block-wise transferable query     & \xmark  & \cmark  & \cmark \\
    \bottomrule
  \end{tabular}
\end{table}

\subsection{Response Caching}\label{sec:doc:caching}

Caching of CoAP responses can bolster packet loss in lossy, constrained networks, but must meet three challenges for an efficient utilization:
\one Consistency of the CoAP \emph{cache key} equivalent DNS queries.
This key is used to determine the existence of cached response copies.
\two Alignment of the CoAP response caching time with DNS record times to live (TTLs) such that a DoC~client does not receive outdated content nor triggers requests too early.
\three Leveraging of CoAP cache validation to reduce the number of large DNS response transmissions on cache timeout.

\paragraph{Consistent cache keys.}
When using CoAP FETCH or GET, the original DNS message becomes part of the cache key, either because the key includes the payload (FETCH) or the URI (GET).
Since all DNS messages carry an ID in its header, which may differ for multiple queries of the same resource record or hostname, we propose to set this ID to 0 for either encrypted CoAP mode.
This yields a deterministic wire format without introducing additional state at the client side or coordinating this ID between multiple DoC~clients.
DTLS and OSCORE provide sufficient defense against spoofing attacks for predictable DNS IDs.

\paragraph{Aligning expiration timers.}
TTLs in DNS responses describe how long a resource record should stay in a DNS cache.
TTLs are actively updated by DNS caches to reflect the remaining cache lifetime.
When embedded in CoAP messages, however, TTLs are opaque since DNS responses are treated like any other CoAP payload.
DoC~clients or proxies thus face two cases that impact the protocol efficiency when cached responses are retrieved.
\one DNS TTLs are expired, although CoAP caches consider them as valid.
This leads to outdated name resolutions at DoC~clients.
\two Cached CoAP responses are expired, while the DNS TTLs are still valid.
This leads to a reduced cache utilization and unnecessary network overhead.

\newcommand*\circled[1]{\protect\tikz[baseline=(char.base)]{\protect\node[shape=circle, draw, inner sep=1pt, font=\normalfont\small] (char) {#1};}}  

\begin{figure}
  \centering
  \includegraphics[width=\columnwidth]{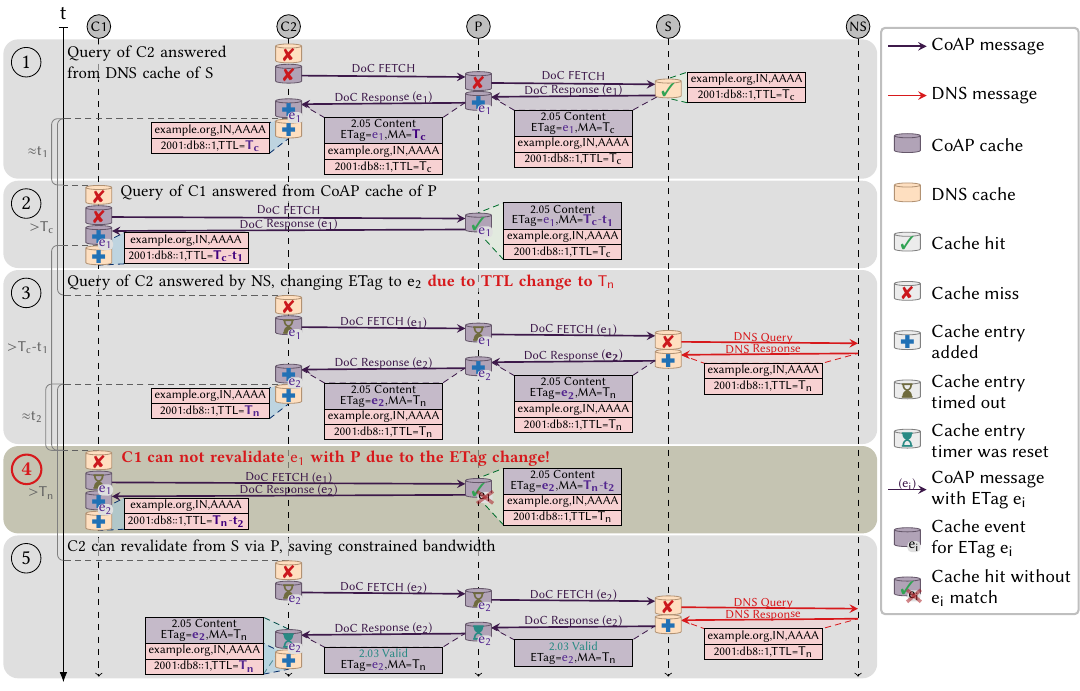}
  \setlength{\abovecaptionskip}{-1em}
  \setlength{\belowcaptionskip}{-1em}
  \caption{A name resolution using the \emph{DoH-like} caching scheme in DoC\@. DoC clients C1 and C2 query \texttt{AAAA} records for \texttt{example.org} from a DoC server S via a DoC-agnostic CoAP proxy P \circled{1} \circled{2}. Only when a DNS cache miss at S occurs, the DNS name server NS is queried \circled{3}. A cache hit at P for a query from C2 adapts the TTL, causing cache re-validation to fail and to resend all information \circled{4}. Only C1 can leverage cache validation immediately just after the Max-Age (MA) update \circled{5}.}%
  \label{fig:caching}
\end{figure}

\paragraph{Leveraging cache validation.}
A CoAP server may include an entity-tag~(ETag) in a response to differentiate between response representations.
Then, client or proxy caches use ETags to query for the validity of timed out cache entries.
If a cached object is still valid, the server transmits a \emph{small} confirmation message using the 2.03 response code to reduce the network load.
Whenever the payload of the entry, \ie the ETag changed, the server transmits the full response.
For our evaluation, we consider two approaches to align the DNS record lifetimes with the CoAP response caching and validation model:
\emph{(1) DoH-like}, which is based on the caching recommendations for DoH~\cite{RFC-8484} as baseline and \emph{(2) EOL TTLs}, our improvement.
\emph{EOL TTLs} leverages CoAP cache validation by setting all TTLs to 0, \ie their End of Live~(EOL).

\emph{Option 1: DoH-like.}
This approach strictly follows RFC 8484~\cite{RFC-8484}.
An example is illustrated in \cref{fig:caching}.
A DoC server sets the Max-Age option of CoAP to the minimum TTL of encapsulated DNS resource records.
The Max-Age value decreases on intermediate CoAP caches.
DoC clients that receive a cached response use the altered Max-Age to reduce TTLs of included resource records (\eg \circled{2} and \circled{4} in \cref{fig:caching}).
A drawback of this approach is that changing DNS TTLs, either due to DNS caches or DNS operators, changes the CoAP payload and thereby affects the ETag generation (see \circled{3}).
This results in failing cache re-validations and requires a full DNS response, even though only the TTL changed (see \circled{4}).
In common DoH deployments such overhead is of not much concern but it reduces performance in the IoT significantly (see \cref{sec:eval-caching} for details).

\emph{Option 2: EOL TTLs.}
Steps \circled{3} and \circled{4} in \cref{fig:caching} illustrate the problem of the \emph{DoH-like} approach. Due to changing TTL values in the DNS cache infrastructure, the cache validation model of CoAP can fail.
 \emph{EOL TTLs} improves this situation for increasing success rates of cache re-validations.
In detail, we propose that a DoC~server sets the Max-Age CoAP option to the minimum TTL of the resource records in the DNS response and rewrites all DNS TTLs to 0.

These modifications always ensure identical ETag values for the same resource record set.
During name resolution, such responses may be stored at intermediary caches, \eg on a proxy or on a client.
For later requests that result in cache hits, Max-Age values are adjusted according to the CoAP freshness model.
When a DoC client receives a response, it copies the CoAP Max-Age into the DNS resource records to restore the correctly decremented TTL values before placing them in the local DNS system cache.
Requests that hit stale cache entries trigger a cache re-validation.
If only TTLs changed, then the DoC server validates the cache entries, and encodes the new TTL in the Max-Age~option, leveraging the cache validation already in step \circled{4} of \cref{fig:caching}, which \emph{DoH-like} was only able to obtain in step \circled{5}.

\subsection{Security Modes}

\begin{figure}
  \centering
  \captionsetup[subfloat]{captionskip=-0.3em}
  \setlength{\abovecaptionskip}{-0.0em}
  \setlength{\belowcaptionskip}{-1em}
  \subfloat[Transport Layer Security (DTLS)]{%
    \label{fig:security-mode-transport}
    \centering
    \includegraphics{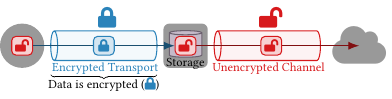}
  }
  \subfloat[Content Object Security (OSCORE)]{%
    \label{fig:security-mode-object}
    \centering
    \includegraphics{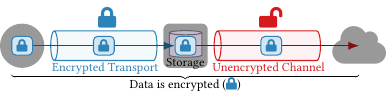}
  }
  \caption{The security modes available with CoAP\@.}%
  \label{fig:security-mode}
\end{figure}

The use of CoAP as transport for DNS enables two security modes to encrypt name resolutions:
\one Transport layer security (see \cref{fig:security-mode-transport}) or \two content object security (see \cref{fig:security-mode-object}).
Both modes can also be used in combination.

\paragraph{Transport Layer Security.}
CoAP over DTLS~(CoAPS)~\cite{RFC-6347} exhibits similar protocol behavior and security guarantees (\ie confidentiality and integrity) as TLS~\cite{RFC-8446}, and further contributes a modified record layer that tolerates message reordering and packet loss.
One key advantage of this mode compared to DoDTLS without CoAP is the block-wise transfer to fragment large DNS queries and responses into smaller CoAP messages.
This resolves a major limitation of DoDTLS\@:
The datagram-based DTLS cannot deal with packets that exceed the PMTU but relies on negotiating a maximum response size during name resolution~\cite{RFC-8094}.

DTLS has two drawbacks impairing DoC\@:
First, without cross-layering, each CoAP retransmission needs to be re-encrypted and re-authenticated before delivery---retransmissions of the encrypted datagram may be rejected by the duplicate detection of the peer.
Second, CoAP proxies and intermediary caches must be included in the trust relationship to process CoAP messages.

\paragraph{Content Object Security.}
OSCORE~\cite{RFC-8613} is a CoAP protocol extension that addresses the drawbacks of DTLS\@.
Instead of securing transport sessions between endpoint pairs, it provides integrity, authenticity, and confidentiality on an object level by protecting \emph{entire} CoAP messages.
OSCORE transforms the original CoAP message into an authenticated and encrypted CBOR Object Signing and Encryption~(COSE)~\cite{RFC-8152} object, and encapsulates it as a CoAP option in an outer, newly created CoAP header, which only exposes the request-response mapping token, message-layer ID and the unprotected parts of the OSCORE option.

To mitigate mismatch and replay attacks, OSCORE constructs a strong message binding between requests and corresponding responses with the use of identical identifiers in their authenticated components. These persist over retransmissions.
OSCORE also reduces computational effort for encryption and authentication compared to CoAP over DTLS, since the protected messages are stored in the CoAP retransmission buffer.

To enable caching on untrusted nodes, a protocol add-on for OSCORE is currently discussed~\cite{draft-amsuess-core-cachable-oscore}, thus ensuring end-to-end security via third-party gateways.
Likewise, there is a proposal to allow for protected group requests and responses for one-to-many communication~\cite{draft-ietf-core-oscore-groupcomm}.

OSCORE initially relies on pre-shared keys or preconfigured certificates.
DTLS provides a built-in key exchange protocol to establish temporary session keys between two endpoints.
This enables perfect forward secrecy: Leaked keys cannot be used to decrypt past correspondences.
A lightweight authenticated key exchange for OSCORE is under development, though: Ephemeral Diffie-Hellman over COSE~(EDHOC)~\cite{draft-ietf-lake-edhoc}.


%% file: tex/evaluation-base.tex
\section{Comparison of Low-Power DNS Transports}\label{sec:eval-base}
In this section, we evaluate memory usage, packet sizes, and resolution times of DoC and compare with DNS over UDP and DNS over DTLS in different communication setups.
Our DoC configurations include the unencrypted use, CoAPS, and OSCORE, using the FETCH, GET, and POST methods.
Last, we provide a numerical comparison with DoQ based on previous work.
For sake of brevity we excluded an evaluation of block-wise transfer and the packet size analysis for GET and POST here.
Both can be found in \cref{sec:eval:pkt-size:coap}.

\subsection{Setup}\label{sec:eval:setup}
\paragraph{Hardware and software platform.}
We conduct our experiments in the FIT IoT-LAB testbed~\cite{abfhm-filso-15}, which supports a variety of IoT hardware environments.
We choose nodes from the \emph{Grenoble} site, since the physical stretch makes it a good candidate for multi-hop measurements.
Our platform features a Cortex-M3 MCU with 64~kBytes of RAM, 512~kBytes of ROM~\cite{st-stm32f103rey-18}, and an IEEE 802.15.4 radio~\cite{microchip-at86rf231-09}.
The radio is configured to automatically handle link layer retransmissions and acknowledgments.

As the base of our experiments, we use RIOT (\texttt{2022.07}), which includes DNS over UDP\@.
We provide extensions to support DNS over DTLS and DNS over CoAP\@.
For details about the implementations, we refer to Appendix~\ref{sec:impl}.
We stress-test each of the deployments by using asynchronous protocol features that allow for concurrently pending queries on a device.
We modify a few RIOT configuration parameters to accommodate the number of queries in the air, specifically internal queue sizes to hold multiple packets.

Since DTLSv1.2~\cite{RFC-6347} still enjoys a wider deployment at the time of evaluation, we choose that version in our evaluations over 1.3~\cite{RFC-9147}.
For consistent measurements, we pre-initialize DTLS sessions and OSCORE replay windows on all endpoints before starting experiments.
To prevent side effects such as lost requests or prolonged timeouts due to session re-initialization, we increase both the DTLS session timeout and the OSCORE replay window size.
This proved useful when measuring the protocol effects during long experiment runs.
To make the 6LoWPAN implementations of RIOT (clients) and Linux (resolver) more comparable, we deactivate the stateful address compression and set the traffic class and flow label IPv6 header fields to \texttt{0}, so they are elided.

In all experiment runs, we measure the actual name resolutions within the IoT network, and exclude the resolution times to external DNS servers.

\paragraph{Topology description.}
We construct a topology with two wireless hops as in \cref{fig:architecture} for two DNS clients communicating with a DNS recursive resolver via a forwarder and a border router.
The forwarder is either configured as an opaque IPv6 router, or as a CoAP forward proxy with caching capabilities.
The border router node is of the same hardware as the DNS clients and the forwarder.
It further connects to the host machine of the DNS resolver via Ethernet that is encapsulated in a TCP-tunneled UART connection.
The DNS resolver is a simple Python implementation that uses standard libraries, such as \emph{dnspython} and \emph{aiocoap}, and runs on the SSH frontend.
The routes in the wireless domain are constructed using the RPL routing protocol~\cite{RFC-6553}.

\paragraph{Protocol settings.}
We evaluate the following DNS transports (short names in parentheses):
\one DNS over UDP (\textbf{UDP}),
\two DNS over DTLS (\textbf{DTLSv1.2}),
\three DNS over unencrypted CoAP (\textbf{CoAP}),
\four DNS over CoAP over DTLS (\textbf{CoAPSv1.2}), and
\five DNS over OSCORE (\textbf{OSCORE}).
We assess CoAP and CoAPSv1.2 with the FETCH, GET, and POST methods, for OSCORE we use only FETCH since our DNS over OSCORE implementation does not support GET due to its complexity.
With DTLSv1.2 we use the \texttt{AES-128-CCM-8} cipher suite~\cite{RFC-6655} and with OSCORE the \texttt{AES-CCM-16-64-128} cipher mode~\cite{RFC-8152}, as these are the most comparable options.  
Both pre-shared key lengths are 9 bytes.

\begin{figure}
  \centering
  \captionsetup[subfloat]{captionskip=-0.2em}
  \setlength{\abovecaptionskip}{-0em}
  \setlength{\belowcaptionskip}{-1.5em}
  \subfloat[ROM]{
    \includegraphics{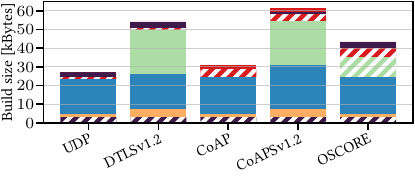}
  }
  \subfloat[RAM]{
    \adjustbox{clip=True, trim=1.7em 0 0 0}{%
      \includegraphics{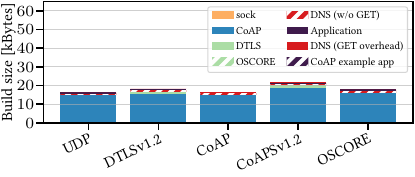}%
    }
  }
  \caption{%
    Memory consumption of each DNS transport with existing CoAP example application.
  }%
  \label{fig:mem-w-app}
\end{figure}

\paragraph{Communication setup.}
We query \texttt{A} and \texttt{AAAA} records in separate runs for 50 names of length 24 characters each, emulating real world data by choosing the median name length of the empirically analyzed names in \cref{sec:dns-empiric}.
The name encodes an identifier to track query-response pairs over the different DNS transports, even if the transaction ID is set to 0.
The query rate is Poisson-distributed with $\lambda = 5$ queries/s to generate noticeable load on the medium.
The recursive resolver is mocked up to generate the desired responses.
For the CoAP-based transport, the requested DNS resource is \texttt{/dns}.
All runs are repeated 10 times.

\subsection{Memory Consumption}\label{sec:eval:mem-size}

We first inspect the memory consumption on our target platform for the DNS requester application of each transport.
For better comparability, we also add the standard \emph{gCoAP} example of RIOT, providing both server and client functionality, to account for a CoAP application already present on the device.
Since the asynchronous request contexts consume a disproportionate amount of RAM compared to the core functions of each protocol, we limit the maximum number of these contexts to one.
As our software platform does not use any dynamic memory allocation, we do not consider heap allocation.
For measuring ROM requirements, we sum up the respective object sizes in the \verb$.text$ and \verb$.data$ sections of the RIOT image and for the RAM requirements the \verb$.data$ and \verb$.bss$ sections.
See Appendix~\ref{sec:eval:details} for details.

\cref{fig:mem-w-app} displays the RAM and ROM consumption for the selected protocols including applications.
The encrypted transports add a considerable amount of ROM---about 24 kBytes in the case of DTLS and about 11 kBytes in the case of OSCORE---and in the case of DTLS also about 1.5 kBytes of RAM\@.
Notably, the DTLS part of the firmware expects more than double the memory space of the OSCORE part.
This is due to DTLS requiring its own message layer, as well as asymmetric cryptography, to establish a handshake, which is not present in OSCORE\@.

GET support adds about 2 kBytes of ROM and 173 bytes of RAM to the overall size.
About 1 kByte of this ROM contributes the URI template processor.
The remainder relates to the different message handling required for the GET request, while the Content-Format option is elided.

These numbers show that for unencrypted transport, UDP remains the clear choice when it comes to memory efficiency.
For encrypted DNS communication, DTLS is the most efficient transport solution, with OSCORE being a close second.
If CoAP is already present for the application, OSCORE is the most efficient encrypted transport.

The comparably young DNS part for DoC has definite potential for optimization.
Currently, it includes parts of the parsing and handling of certain CoAP options and is with around 4 kBytes significantly larger than the other DNS transport implementations.
This header handling should be moved to the CoAP part of the firmware in the future to remove possible code duplication with application code.

\begin{figure}
  \centering
  \setlength{\belowcaptionskip}{-1em}
  \includegraphics[width=\textwidth]{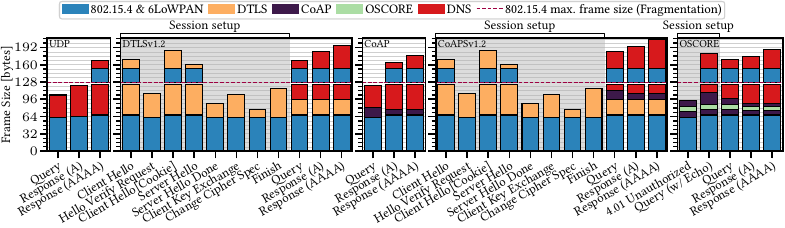}
  \caption{Maximum link layer packet sizes for each transport when resolving a name with a length of 24 characters for a single record (\texttt{A} or \texttt{AAAA} respectively).}%
  \label{fig:eval-pkt-size}
\end{figure}

\subsection{Packet Sizes}\label{sec:eval:pkt-size}

We now measure the packet sizes of the DNS messages on all transports by capturing the IEEE 802.15.4 frames using the \texttt{sniffer\_aggregator} tool of the FIT IoT-LAB testbed.
\cref{fig:eval-pkt-size} displays the packet dissection for each packet type, segmented per communication layer.
Both IPv6 and UDP headers are compressed within the 6LoWPAN header, which we group with the IEEE 802.15.4 MAC header for simplicity.
The maximum PDU of IEEE 802.15.4 is marked in each plot by a red dashed line.
6LoWPAN fragments larger IEEE 802.15.4 frames, producing additional MAC and 6LoWPAN headers for each generated fragment.
We represent each additional fragment with its headers above the red marker line.

We see three distinct sizes of DNS messages in our experiments.
DNS queries requesting either an \texttt{A} or \texttt{AAAA} record from the DNS resolver.
These queries are identical in size and only differ in their query types (\texttt{A} vs.~\texttt{AAAA}).
Respective responses contain either an \texttt{A} or \texttt{AAAA} record, which vary in size due to the IP address lengths.

\cref{fig:eval-pkt-size} further includes packet sizes of the DTLSv1.2 handshake and the OSCORE replay window initialization.
We observe that DTLS---both with DTLSv1.2 and CoAPSv1.2---is at a disadvantage as the handshake messages alone already cause multiple fragmented datagrams and multiply the likelihood for packet loss during the session establishment.

DNS queries are \texttt{base64}-encoded within the GET method.
This inflates requests to a size that is approximately 1.5 times larger than binary FETCH or POST queries.
As such, with either CoAP-based transport a DNS query using GET will be fragmented for the median name length.
Likewise, when \texttt{AAAA} records are requested the response will be fragmented.
For CoAPSv1.2, little room is left for the DNS message itself in either message format before reaching the fragmentation limit.
The same, however, is also true for OSCORE, if the Echo option required for the replay window initialization is carried in the request.
\Cref{sec:eval:pkt-size:coap} adds supplementary dissections for queries using GET\@.

Overall, for unencrypted transmission, UDP is the preferred transport for mitigating fragmentation.
For encrypted usage, OSCORE is the preferred method.
CoAP pack\-ets can multiplex message formats with the Content-Format option.
A new compressed DNS messages format could thus help to mitigate fragmentation for CoAP-based transports.

\subsection{Name Resolution Times}\label{sec:eval-nrt}
\begin{figure}
  \centering
  \captionsetup[subfloat]{captionskip=-0.1em}
  \setlength{\abovecaptionskip}{-0.1em}
  \setlength{\belowcaptionskip}{-1em}
  \subfloat[A record]{%
    \includegraphics{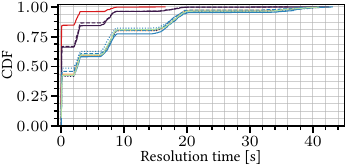}%
  }
  \subfloat[AAAA record]{%
    \adjustbox{clip=True, trim=2.3em 0 0 0}{%
      \includegraphics{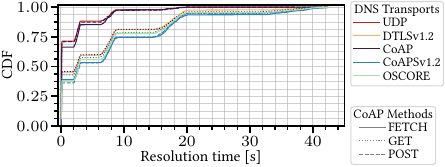}
    }
  }
	\caption{Resolution times for 50 queries (Poisson distributed with $\lambda = 5$ queries/s).}%
  \label{fig:load-cdf}
\end{figure}

Next, we evaluate the name resolution times for each protocol.
For this, we measure the time from issuing the query by the DNS client until the IP address is parsed in the response.

\cref{fig:load-cdf} summarizes the distributions of resolution times for our protocols.
We observe that the different transports form distinct groups in their temporal distributions due to the different packet sizes and the resulting 6LoWPAN fragmentation.
For UDP requesting an \texttt{A} record no packet is fragmented and names resolve fastest.
85\% of the queries resolve in less than 250~ms, all complete within 20~s.
Requesting an \texttt{AAAA} record, though, plain UDP compares to unencrypted CoAP with the FETCH or POST method.
The query is not fragmented, but the response is.
For these transports and methods, only 65--70\% of the names resolve below 250~ms, but 99\% of names resolving within 20~s.
The last group consists of those transports and methods, for which both queries and responses fragment.
Unencrypted CoAP with GET, as well as DTLSv1.2, CoAPSv1.2, and OSCORE perform all within approximately 7\% from each other, with 42--49\% of \texttt{A} records and 37--45\% of \texttt{AAAA} records resolve below 250~ms.
All require at most 41--44~s to resolve 99\% of names.
These long resolution times are due to the CoAP retransmission algorithm and of no concern when considering typical IoT duty cycles, such as the ones employed by LoRaWAN or IEEE 802.15.4e.

As resolution time closely relates to packet size and the number of fragments, the same conclusions as in \cref{sec:eval:pkt-size} can be drawn.

\subsection{Comparison with DNS over QUIC}\label{sec:eval:quic}

\begin{figure*}
  \setlength{\belowcaptionskip}{-1em}
  \begin{minipage}{52mm}
    \centering
    \includegraphics{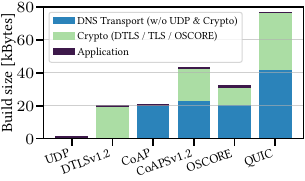}
    \caption{%
      Code sizes of UDP-based DNS transports.
    }%
    \label{fig:mem-quic}
  \end{minipage}
  \hfill
  \begin{minipage}{84mm}
    \centering
    \captionsetup[subfloat]{captionskip=-0.0em}
    \subfloat[0-RTT packet]{%
      \includegraphics{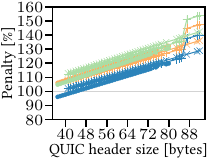}%
      \label{fig:eval-pkt-size-quic:0rtt}
    }%
    \subfloat[1-RTT packet]{%
      \adjustbox{clip=True, trim=2.1em 0 0 0}{%
        \includegraphics{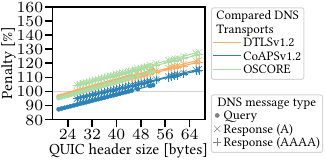}%
        \label{fig:eval-pkt-size-quic:1rtt}
      }%
    }%
    \caption{Relative amount of additional link layer data required by DNS over QUIC compared to other DNS transports when resolving a name with a length of 24 characters for a single record (\texttt{A} or \texttt{AAAA}).}%
    \label{fig:eval-pkt-size-quic}
  \end{minipage}
\end{figure*}

DNS over QUIC~\cite{RFC-9250} could be a lightweight alternative to DNS over CoAP because DNS over QUIC is based on UDP, an IoT-friendly transport.
We now compare DNS over QUIC with DNS over CoAPS, DTLS, and OSCORE\@.
To analyze the code size of implementations, we base our comparison on prior work
about QUIC in the IoT~\cite{e-tsitq-20}.
To analyze message sizes, we conduct a numerical evaluation.

\paragraph{Code size.}
Our point of reference is \emph{Quant}~\cite{e-tsitq-20}, a QUIC implementation on top of RIOT\@.
To account for differences in implementations that are not protocol-specific, we intentionally omit the UDP layer and the \emph{sock} part of RIOT, because Quant accesses the networking modules of RIOT differently than our DNS client implementation.
For further details, we refer to Appendix~\ref{sec:eval:details}.

\cref{fig:mem-quic} shows the amount of data \verb!.text! and \verb!.data! sections of the generated binaries require.
QUIC, including TLS, uses nearly double the ROM as any of the common IoT transports.
It is worth noting that Quant includes only the QUIC~client part while our implementations include both CoAP~client and server code.
Further optimizations proposed in~\cite{e-tsitq-20} can only save $\approx$20~kBytes, which would require DNS over QUIC to use more ROM compared to DNS over CoAP\@.

\paragraph{Packet size.}
QUIC has variable header lengths for two reasons.
First, different types of handshakes, 0-RTT, which complete without additional round trips, and 1-RTT, which require an additional round trip, lead to different header types.
Second, header fields (\eg connection IDs) can be of variable sizes.
To assess packet sizes realistically, we conduct a best and worst case analysis for both 0-RTT and 1-RTT handshakes.

\cref{fig:eval-pkt-size-quic} shows the relative amount of link layer packet sizes DNS over QUIC would require to resolve a 24~chars name for a single \texttt{A} and \texttt{AAAA} record, compared to the other DNS transports.
In the best case, \ie 1-RTT handshakes with small headers, DNS over QUIC is comparable to DNS over CoAP, but in the majority of cases DNS over CoAPS, DTLS, and OSCORE outperform DNS over QUIC (see \cref{fig:eval-pkt-size-quic:1rtt}).
In case of 0-RTT QUIC handshakes, efficiency of DNS over QUIC decreases even more (\cref{fig:eval-pkt-size-quic:0rtt}).
Requesting an IPv6~address in max header scenarios will trigger fragmentation into 3~fragments to carry the \texttt{AAAA} response over QUIC\@.

\paragraph{Summary.}
DNS over QUIC is less suitable for the IoT than DNS over CoAP\@.
The code size is larger and packet sizes will require very use-case specific tweaking to selected header fields.


%% file: tex/evaluation-caching.tex
\section{Evaluation of Caching for DoC}\label{sec:eval-caching}
In this section, we perform a comparative assessment of caching as introduced in \cref{sec:doc} using a multihop network.

\subsection{Setup}
We base our evaluation on the setup described in \cref{sec:eval:setup}, except that clients now query 50 \texttt{AAAA} records of only eight distinct names to showcase the cache utilization.



We compare our two approaches, \emph{DoH-like} and \emph{EOL TTLs} (see \cref{sec:doc:caching}) with three levels of caching:
\one a DNS cache at each client,
\two a CoAP cache at each client,
and \three a CoAP cache at the forwarder, which runs as CoAP forward proxy (see \cref{fig:architecture}).
When no cache at the forwarder and thus no forward proxy, we call these the \emph{opaque forwarder} scenarios.

To focus on the impact of caching, we only evaluate unencrypted CoAP to reduce the packet size overhead of encryption and unrelated side effects stemming from that.
The DNS resolver returns four \texttt{AAAA} records for each name query.
This causes 6LoWPAN fragmentation with three fragments for the responses.
All four re\-cords use the same TTL, uniformly picked between 2 and~8~s to introduce quick cache renewals.

\subsection{Link Utilization}\label{sec:eval-caching:lu}
\begin{figure*}[t]
  \captionsetup[subfloat]{captionskip=-0.1em}
  \setlength{\belowcaptionskip}{-1em}
  \subfloat[Opaque forwarder]{%
    \centering
    \includegraphics{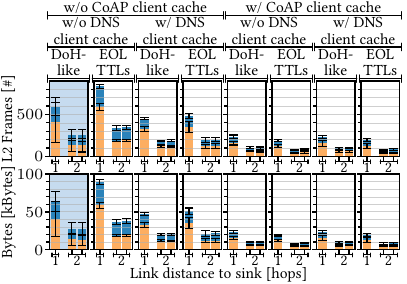}%
    \label{fig:eval-caching-lu:proxied0}
  }
  \hspace{0em}
  \subfloat[With caching CoAP proxy]{%
    \centering
    \includegraphics{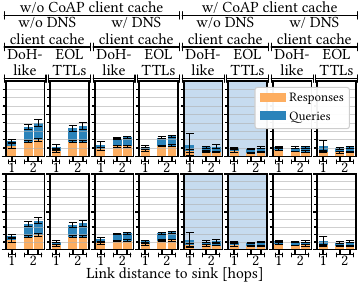}%
    \label{fig:eval-caching-lu:proxied1}
  }
  \caption{Link utilization for four \texttt{AAAA} record queries (Poisson-distributed with $\lambda = 5$ queries\,/\,s) for different caching solutions with CoAP FETCH, comparing DoH-like and EOL TTLs caching approaches. The scenarios highlighted in blue are evaluated in detail in \cref{sec:eval-caching:rch}.}%
  \label{fig:eval-caching-lu}
\end{figure*}

We show the influence of the FETCH method on the links for different caching scenarios in \cref{fig:eval-caching-lu}.
The bars with 1-hop link distance to the sink represents the link utilization between proxy and border router, the bars with distance 2 the utilization between each client and the proxy in \cref{fig:architecture}.

Providing a CoAP cache decreases load on all links, and using the \emph{EOL TTLs} approach instead of \emph{DoH-like} further decreases the load.
Using DNS caching at the clients gives only little advantages.
CoAP caching leads to 50\% less link utilization in any configuration.
Depending on which node (proxy or client) provides the CoAP cache, the upstream link directly benefits from the cache.
Due to the higher amount of messages on the bottleneck link between proxy and border router, the caching can unfold its full potential, visible at that link in the \emph{EOL TTLs} cases in \cref{fig:eval-caching-lu:proxied1}.
Here, $\approx$50~frames less and, depending on the cache configuration, 5--10~kBytes less need to be exchanged.
Small advantages can also be observed at the client-proxy links.

\subsection{Transport Retransmissions, Cache Hits}\label{sec:eval-caching:rch}
We now quantify the link stress that clients generate due to corrective actions.
We track both the timestamp of each event at which a client initiates a CoAP transmission and the timestamp for cache hits including re-validations of stale entries at the clients and the proxy.
For all events, we calculate the time offset to the start time of the respective DNS query.
We focus on the blue scenarios of \cref{fig:eval-caching-lu}, because they are the most interesting, and consider GET and POST in addition to FETCH\@.

\begin{figure}[t]
  \centering
  \setlength{\belowcaptionskip}{-1em}
  \includegraphics{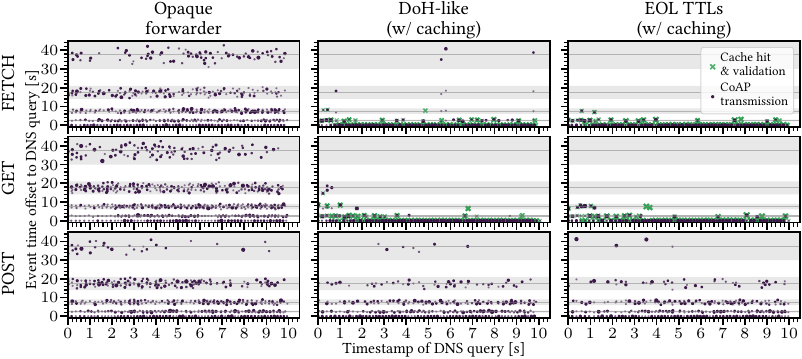}
	\caption{CoAP events of message (re-)transmissions at the client compared to the time of the initial DNS query. Retransmissions follow an exponential back-off and scatter within specific (gray) areas. Size and opacity of each marker represent the multiplicity of events in the same area.}
  \label{fig:eval-proxy-trans}
\end{figure}

The original requests have a negligible time offset in the range of microseconds, and since retransmissions follow a random exponential back-off mechanism~\cite{RFC-7252}, their time offsets scatter within specific regions (gray areas in \cref{fig:eval-proxy-trans}).

FETCH in combination with \emph{EOL TTLs} provides the best performance (see \cref{fig:eval-proxy-trans}).
Cache hits are able to complete requests without requiring more than one retransmission for most of the DNS queries.
POST requests do not utilize response caches, which degrades their performance to the level of the \emph{opaque forwarder}.

In the \emph{opaque forwarder} scenario, we observe about 50\% more retransmissions for both GET and FETCH compared to any of the \emph{caching approaches}.
Using GET, the retransmissions in the third and fourth iteration even increase by 7\% compared to POST and FETCH\@. This is a result of 6LoWPAN fragmenting the query (see \cref{sec:eval:pkt-size}).

%% file: tex/discussion.tex
\section{Discussion and Optimization Potentials}\label{sec:discussion}

With the core DoC protocol at hand, impactful opportunities for security and privacy open. In addition, a series of potential optimizations and protocol enhancements become attainable, which---along with open questions and limitations---we discuss in the following.

\paragraph{Implications for security and privacy in the IoT\@.}
DoC introduces with CoAPS a new and efficient baseline for privacy and integrity of DNS queries. To this end, our protocol proposal mainly translates DNS over HTTPS to the constrained IoT\@. More importantly, the CoAP extension OSCORE allows for securing the DNS query messages independent of the transport. This approach to content object security enables name resolution with secure messages that are cacheable and can traverse a gateway without requiring a trust relation for transcoding.
This implies that IoT devices can soften their trust relation with gateways in the future.

\paragraph{How to reduce the DNS packet overhead?}
DNS messages account for the largest parts of the packets in DoC\@.
Hence, DNS compression schemes beyond the generic methods of 6LoWPAN~\cite{RFC-4944} or SCHC~\cite{RFC-8724} promise enhanced efficiency as we concluded in \cref{sec:eval:pkt-size}.

One obvious improvement arrives when operators prefer shorter host names in constrained IoT scenarios.
Query name minimization~\cite{RFC-7816,k-mdritp-22}, as applied by the DNS~infrastructure, is not an option, as it relies on the queried server being the authoritative name server of the remaining name part.
However, superfluous DNS header fields may be compressed.
Klauck and Kirsche~\cite{kk2013ednsmc} proposed compression for mDNS/DNS-SD messages for 6LoWPAN, but their approach focuses on compatibility with the DNS wire format.
DoC offers the opportunity to use different message formats via its use of a new Content-Format.
Specifically, the Concise Binary Object Representation~(CBOR)~\cite{RFC-8949} offers a standardized, structured, and space-efficient encoding.

In addition, CoAP messages carry a transactional context that matches a reply to its request. Exploiting this, we argue for the following practices to reduce packet overhead.
A DoC question could be encoded as a CBOR array, containing up to three entries: the name (as text string), an optional record type (as unsigned integer), and an optional record class (as unsigned integer).
If record type and class are elided, DoC implies \texttt{AAAA} and \texttt{IN}, respectively.
A DoC response can be matched to the request. Hence, the encoding could use only one CBOR array,
which contains the DNS answer section. This could be nested for several, separate answer sections. An answer section includes a TTL, a name, and an optional type specifier using the space-efficient encoding of CBOR\@.
For an answer with two arrays, DoC additionally identifies the question section (formatted as specified above).

In our evaluation, we could verify that the wire-format of an \texttt{AAAA} response packet compresses from 70 bytes (see \cref{fig:eval-pkt-size}) down to 24 bytes---a reduction by 66\%.
Further suffix and prefix compression, as well as referencing redundant values, can be provided with Packed CBOR~\cite{draft-ietf-cbor-packed}.
The CBOR working group of the IETF currently discusses such a format~\cite{draft-lenders-dns-cbor}.

\paragraph{How to protect the integrity of the DNS TTLs?}
DoC depends on the CoAP Max-Age option to track elapsed caching time, which a DoC client then uses to decrement DNS TTLs.
The integrity of the Max-Age option, however, cannot be guaranteed, because it is altered on---potentially untrusted---intermediaries.
An adversary with malicious intent, or a faulty proxy behavior may impair TTLs on the client by using incorrect Max-Age values.

For \emph{EOL TTLs}, a potential mitigation is to include a second Max-Age value that is protected by OSCORE\@.
A DoC client compares both Max-Age values, deduces inconsistent modifications, \eg larger values than the original TTLs, and discards the response when the consistency check fails.
For the \emph{DoH-like} caching scheme, responses include the original TTLs, which can be used to perform consistency checks instead of including an additional Max-Age value.
This approach mitigates the use of outdated DNS records, but still allows for unauthorized reduction of TTLs, which affects the caching~performance.

\paragraph{How to support DNS load balancing and cache re-validation?}
A na\"{\i}ve ETag generation calculates a hash over the CoAP message payload to identify a specific DNS response.
Common DNS behavior challenges this.
In addition to changing TTLs, DNS resolvers often rearrange resource records within responses for load balancing reasons.
This modifies the binary representation of DNS messages, and thus their resulting ETag values.
One approach to support load balancing without altering the message is to sort incoming records at the DoC server and randomize records at the DoC client.

\paragraph{How to utilize OSCORE group communication in DNS?}
An advantage of OSCORE over other encrypted DNS transports, including DoH and DoT, is the support of group communication~\cite{draft-ietf-core-oscore-groupcomm}.
This qualifies OSCORE for encrypted DNS-based service discovery utilizing multicast DNS\@.
On the downside, multicast can be very energy intensive with larger impact especially in constrained networks.
Future work should analyze DNS over Group OSCORE and carefully weigh benefits and drawbacks.

\paragraph{Are our insights limited to CoAP?}
No.
Our analysis of domain names may guide protocol engineering for DNS resolution in constrained networking scenarios in general.
Some protocols may introduce additional challenges, though, such as the asynchronous communication model in MQTT\@.

\paragraph{Why DNS over CoAP at all?}
CoAP has several advantages compared to DTLS to secure DNS resolution.
First, from the system level, many applications already rely on CoAP\@.
Reusing software components that are part of the system stack reduces memory requirements (see \cref{sec:eval-base}).
Second, resolving names on top of an application layer protocol that is not dedicated to name resolution mitigates common censorship approaches.
CoAP obfuscates the embedded service---similar to DNS over HTTPS running on standard HTTPS TCP/443 instead of TLS running on the DNS-specific well-known port TCP/853.
Third, combined with OSCORE, CoAP yields the potential of both secure and more resilient communication via encrypted, en-route caching.
Third, when considering LPWANs, OSCORE object security offers the opportunity for further header compression using SCHC~\cite{RFC-8824}, providing additional space for the encrypted DNS message.
DTLS security cannot achieve this because transport security obfuscates those header fields on which compression is based.

Simply mapping CoAP to DoH (\eg using a CoAP-HTTP proxy) would also introduce drawbacks.
First, a proxy deployment adds extra bytes to CoAP packets.
Second, HTTP primitives do not provide benefits gained from CoAP-specific FETCH (see \cref{sec:eval-base}) and dedicated \emph{EOL TTLs} (see \cref{sec:eval-caching}).


%% file: tex/conclusion.tex
\section{Conclusion and Outlook}\label{sec:conclusion}

The constrained IoT lacks a protocol for privacy-friendly, secure name resolution like DoH provides for the broader Internet.
We presented DNS over CoAP~(DoC), which leverages the rich feature set of the CoAP protocol suite to provide an energy-efficient and end-to-end protected name resolution for constrained networks.
In a comprehensive analysis, we compared DNS over UDP, CoAP, DTLS, CoAP over DTLS, and CoAP with OSCORE in experiments based on full-featured implementations.

Our findings revealed that the performance of name resolution is primarily driven by packet sizes.
While CoAP has a space-efficient protocol encoding, the choice of the request method largely impacts the packet overhead and the additional CoAP features at hand.
 FETCH was identified as the preferred method compared to GET and POST, because FETCH allows for block-wise transfer of queries and for caching of responses.
Correspondingly, OSCORE outperformed CoAPS for protecting DoC as it
 seamlessly integrates with the semantics of CoAP and its header encoding, which reduces packet sizes and memory consumption.
Integrating DNS TTLs with CoAP message lifetimes favors the cache validation model of CoAP and reduces bandwidth demands as well as loss rates further.

Future work shall optimize the coding efficiency by defining a comprehensive compression scheme for DNS messages.
This will enfold impact by reducing fragmentation and by increasing reliability in low-power and lossy regimes.
We will also focus on a DoC integration for mDNS protected by Group OSCORE to enable service discovery.


%% file: tex/acks.tex
\begin{acks}
We would like to thank the anonymous reviewers for their feedback on this paper, as well as the members of the IETF CoRE and DNSOP working groups on their input on DoC.
This work was supported in parts by the \grantsponsor{BMBF}{German Federal Ministry of Education and Research~(BMBF)}{https://www.bmbf.de/} within the projects \grantnum{BMBF}{PIVOT} and \grantnum{BMBF}{PRIMEnet}.
\end{acks}


%% file: tex/appendix_blockwise.tex
\begin{figure}
  \centering
  \setlength{\belowcaptionskip}{-1em}
  \definecolor{block-num}{HTML}{9e0142}
  \subfloat[Block1 for requests]{
    \includegraphics{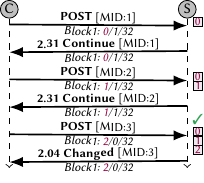}\label{fig:coap-blockwise-block1}%
  }
  \hspace{1em}
  \subfloat[Block2 for responses]{
    \includegraphics{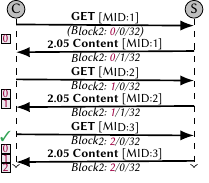}\label{fig:coap-blockwise-block2}%
  }
  \caption{Different block-wise transfers of a 96~bytes body between a client C and a server S using 32~byte blocks in CoAP\@. ${\color{block-num}n}/m/s$ denotes the block number, whether more blocks (or not) can be sent, and the block~size.}%
  \label{fig:coap-blockwise}
\end{figure}

\section{Background on CoAP Block-wise Transfer}\label{sec:block-wise}
CoAP POST and FETCH, which carry a DNS~query in the body of a request, provide an additional advantage in case the DNS query message size exceeds an PDU\@:
The body can be split in the application layer into multiple CoAP messages with block-wise transfer mode to prevent fragmentation on the link layer.
When using the Block1~\cite{RFC-7959} option, a receiver assembles the full message on successful reception of all blocks (see \cref{fig:coap-blockwise-block1}).
In contrast to queries, the Block2 transfer mode~\cite{RFC-7959} allows a client to request a certain block size in the response, but the server may also decide to transfer in blocks proactively, without the Block2 option being present in the initial request (see \cref{fig:coap-blockwise-block2}).


%% file: tex/appendix_impl.tex
\section{Implementation of DoC in RIOT}\label{sec:impl}

In this appendix, we introduce our framework to run DNS queries in the constrained IoT, including implementations of a DoC~prototype, DNS over UDP, and DNS over DTLS\@.
We make use of the IoT operating system RIOT~\cite{bghkl-rosos-18}.

\begin{figure}
  \centering
  \setlength{\belowcaptionskip}{-1em}
  \includegraphics{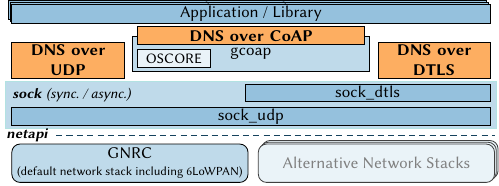}
  \caption{DNS, DoDTLS, and DoC in RIOT\@.}%
  \label{fig:nwstack}
\end{figure}

The network stack of RIOT is visualized in \cref{fig:nwstack} (DNS building blocks in orange).
The \emph{sock} API of RIOT is agnostic to the underlying network stack, such as GNRC, and allows, among other transports, access to UDP and DTLS\@.
Its unified access allows for a seamless composability of network building blocks, eases implementation complexity, and grants a flexible integration of third-party network stacks~\cite{bghkl-rosos-18}.
The existing DNS client in RIOT builds on the \emph{sock} API to interface with the underlying network stack.
Message-related operations to compose DNS queries and parse DNS responses follow a modular and reusable design.
For code simplicity, this client uses the synchronous version of \emph{sock}, so it blocks on requests until a response arrives or a timeout occurs.
The \emph{gCoAP} library provides support for CoAP~\cite{RFC-7252} on top of \emph{sock} as well.
It features all request methods including GET, POST, PUT, and DELETE, but also FETCH, PATCH, and iPATCH~\cite{RFC-8132}.
\emph{gCoAP} also implements proxying capabilities and supports response caches on both proxies and clients~\cite{gasw-cosit-22}.
Block-wise transfer~\cite{RFC-7959} as well as DTLS support are provided.

For the purposes of evaluating DNS over CoAP~(DoC) we extend the stack by the following functionalities.
We extend the DNS over UDP implementation to support asynchronous calls to \emph{sock} for \textbf{non-blocking queries}.
For better comparability with DoC, we support the \textbf{retransmission algorithm of CoAP}~\cite{RFC-7252} for DNS over UDP, \ie 4 retransmissions using an exponential back-off.
We provide a \textbf{DoDTLS client implementation} via \emph{sock} with blocking and asynchronous capabilities and supports Pre-shared Keys~(PSKs) using AES and Elliptic Curve Cryptography~(ECC) via the \emph{TinyDTLS} library.
We implement a \textbf{DoC client} on top of \emph{gCoAP}, which can be configured to use both blocking and asynchronous requests. For OSCORE support, the \emph{libOSCORE} library is used.
We also provide a lightweight \textbf{URI template processor}, which the DoC client can use to marshal the packet format of DNS queries into the URI option of CoAP GET requests.
Both our DoDTLS and DoC implementation reuse the generic interface to compose and parse DNS messages of the DNS over UDP implementation of RIOT\@.


%% file: tex/appendix_riot_config.tex
\section{Details about Experiment Setup}\label{sec:eval:details}

\paragraph{\cref{sec:eval:mem-size}}
We use version \texttt{9-2019-q4-major} of the \textit{GNU ARM Embedded Toolchain}, the recommended toolchain for RIOT \texttt{2022.07}, which includes GCC v9.2.1.  
\cref{tab:riot-ctps} shows the compile-time configuration parameters that we changed from default values.
RIOT ships a tool to dissect the memory usage of a RIOT firmware image from module level down to function and variable level.

We group the modules of RIOT according to the following categories.
\textit{Application} contains all the machine code instructions and state information of the experiment application.
\textit{DNS} contains the code of each DNS over X implementation, including the shared DNS message parser and composer.
As the GET method in DoC adds a significant amount of memory for URI template processing, this is separately shown.
\textit{OSCORE} contains the code of libOSCORE including its dependencies.
\textit{CoAP} contains the code of the gCoAP library and its dependencies, as well as URI parsing.
\textit{sock} contains the code of the sock API implementation for the GNRC network stack, as well as the \emph{sock} implementation for TinyDTLS\@. This was included to account for the different build sizes when using DTLS\@.
\textit{DTLS} contains the code of TinyDTLS including its dependencies.
\textit{CoAP example app} contains the code of the RIOT CoAP example.

\paragraph{\cref{sec:eval:quic}}
We compile Quant (and our DNS over CoAP/DTLS/OSCORE implementations) for RIOT~version \texttt{2022.07} using the \texttt{esp-2021r2-patch3} GCC release by \emph{espressif}.
We used the ESP32~platform because Quant is available on ESP32.

\begin{table}
  \centering
  \caption{Changed compile-time parameters in RIOT 2022.07. An asterisk ($^*$) denotes configuration for the proxy. A plus ($^+$) only applies to the dedicated block-wise runs. All other configurations refer to configurations of the clients.}%
  \label{tab:riot-ctps}
  \footnotesize
  \begin{tabular}{p{11cm}r}
    \toprule
    Compile-time Parameter in RIOT & Value \\
    \midrule
    \texttt{CONFIG\_DNS\_CACHE\_SIZE}           & $8$ \\
    \texttt{CONFIG\_DTLS\_PEER\_MAX}            & $2$ \\
    \texttt{CONFIG\_GCOAP\_DNS\_BLOCK\_SIZE}    & $8^+$ / $16^+$ / $32^+$ / $64^+$ \\
    \texttt{CONFIG\_GCOAP\_PDU\_BUF\_SIZE}      & $228$ \\
    \texttt{CONFIG\_GCOAP\_REQ\_WAITING\_MAX}   & $60$ / $71^*$ \\
    \texttt{CONFIG\_GCOAP\_RESEND\_BUFS\_MAX}   & $60$ / $71^*$ \\
    \texttt{CONFIG\_GNRC\_IPV6\_NIB\_NUMOF}     & $8^*$ \\
    \texttt{CONFIG\_GNRC\_PKTBUF\_SIZE}         & $3072$ \\
    \texttt{CONFIG\_NANOCOAP\_CACHE\_ENTRIES}   & $8$ / $50^*$ \\
    \texttt{CONFIG\_NANOCOAP\_CACHE\_RESPONSE\_SIZE} & $228$ \\
    \texttt{CONFIG\_SOCK\_DODTLS\_RETRIES}      & $4$ \\
    \texttt{CONFIG\_SOCK\_DODTLS\_TIMEOUT\_MS}  & $2000$ \\
    \bottomrule
  \end{tabular}
\end{table}


%% file: tex/appendix_coap_pkt_size.tex
\section{Additional Evaluation Results}\label{sec:eval:pkt-size:coap}

\begin{figure}
  \centering
  \setlength{\abovecaptionskip}{-0.25em}
  \setlength{\belowcaptionskip}{-1em}
  \includegraphics[width=\textwidth]{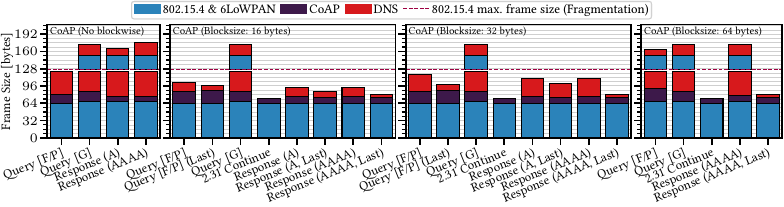}
  \caption{Maximum link layer packet sizes for each transport when resolving a name with a length of 24 characters for a single record (\texttt{A} and \texttt{AAAA} respectively) for different CoAP methods (F = FETCH, G = GET, P = POST) and block sizes. ``Last'' denotes the size of the last block with block-wise transfer.}%
  \label{fig:eval-pkt-size-coap}
\end{figure}

\paragraph{Additional CoAP packet sizes.}
In \cref{sec:eval:pkt-size} we evaluated the packet size, but did not go into detail for the different CoAP message types.
In \cref{fig:eval-pkt-size-coap} we show the packet sizes with block-wise transfer or GET and POST requests for DNS over CoAP\@.

With block-wise transfer, only the payload can be transferred in blocks.
As such, using the GET method, we are unable to send our DNS query in blocks, as it is carried, encoded in \texttt{base64} within the CoAP URI-Query option.
Consequently, the GET request stays the same in all the block-wise transfer modes shown in \cref{fig:eval-pkt-size-coap}.
With block-wise transfer, we are able to reduce the overall packet size enough to drop below the fragmentation line of 6LoWPAN\@.
Compared to fragmentation in 6LoWPAN, CoAP block-wise transfer provides us with a recovery mechanism, so even if a message is lost, we can recover from that on a block-level, so we do not have to send the whole request or response again, in case one single block or fragment gets lost.

For the setup evaluated, a block size of 32 bytes is ideal: 16 bytes makes the blocks smaller and more numerous than necessary and 64 already leads to 6LoWPAN fragmentation.

\begin{figure}
  \centering
  \setlength{\belowcaptionskip}{-1em}
  \captionsetup[subfloat]{captionskip=-0.0em}
  \subfloat[A record]{%
    \includegraphics{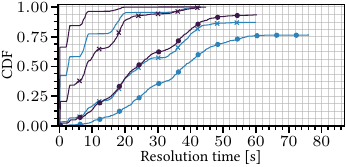}%
  }
  \subfloat[AAAA record]{%
    \adjustbox{clip=True, trim=2.3em 0 0 0}{%
      \includegraphics{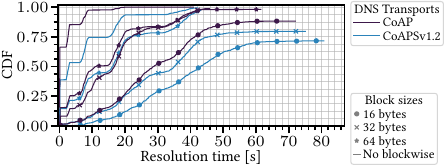}
    }
  }
  \caption{Resolution times for 50 queries using FETCH with block-wise transfer.
  Block size 64 was only used with \texttt{AAAA} records, as DNS responses for \texttt{A} record stay below 64 bytes in size.}\label{fig:load-cdf-blockwise}
\end{figure}

\paragraph{Name resolution times with block-wise transfer.}
To evaluate the name resolution times of block-wise transfers, we used the same communication setup as described in \cref{sec:eval-base} but also statically set the block size for both requests and responses to 16, 32, and 64 bytes, respectively.
Block size 64 was only used with \texttt{AAAA} records, as only the responses for those exceed 64 bytes in the CoAP payload (see \cref{fig:eval-pkt-size-coap}).
We plot the temporal distributions for \texttt{A} and \texttt{AAAA} records with block-wise transfer using FETCH requests for CoAP and CoAPSv1.2 in \cref{fig:load-cdf-blockwise}.
For easier comparison, we include the distributions of \cref{fig:load-cdf}, which do not utilize the block-wise transfer, but we emphasize the increased range of the x-axis.
We observe that the performance decreases with smaller block sizes.
With a block size of 16 bytes, only $\approx$\,90\% and 60--70\% of name resolutions complete in total for CoAP and CoAPSv1.2, respectively.
This is due to congestion emerging in the wireless medium, which increases the probability of packet loss for transfers with higher block counts.
To summarize, block-wise transfer can help mitigate the problem of fragmentation, but leads to a decrease in performance as well, if not properly congestion controlled.


%% file: tex/acronyms.tex
\section{List of Common Acronyms}

{
\setlength{\tabcolsep}{3pt}
\begin{tabular}{rl}
  \bfseries CBOR    & Concise Binary Object Representation \\
  \bfseries CoAP    & Constrained Application Protocol \\
  \bfseries CoAPS   & CoAP over DTLS \\
  \bfseries COSE    & CBOR Object Signing and Encryption \\
  \bfseries DoC     & DNS over CoAP \\
  \bfseries DoDTLS  & DNS over DTLS \\
  \bfseries DoH     & DNS over HTTPS \\
  \bfseries DoT     & DNS over TLS \\
  \bfseries DoQ     & DNS over QUIC \\
  \bfseries DTLS    & Datagram Transport Layer Security \\
  \bfseries OSCORE  & Object Security for Constrained RESTful Environments \\
\end{tabular}
}
